\documentclass[a4paper, 11pt]{article}

\usepackage[top=18truemm,bottom=25truemm,left=25truemm,right=25truemm]{geometry} 
\usepackage{indentfirst} 

\makeatletter 
\def\section{\@startsection {section}{1}{\z@}{-3.5ex plus -1ex minus -.2ex}{2.3 ex plus .2ex}  {\large\bf}}
\makeatother
\usepackage{secdot} 

\makeatletter 
\def\subsection{\@startsection {subsection}{1}{\z@}{-3.5ex plus -1ex minus -.2ex}{2.3 ex plus .2ex}{\small\bf}}
\makeatother

\usepackage{amsmath, amssymb} 

\usepackage[dvipdfmx]{graphicx}
\usepackage{float} 
\graphicspath{{./}} 
\usepackage[labelsep=period, figurename=Fig. ,tablename=Table]{caption} 
\usepackage{subcaption} 
\usepackage{booktabs}
\usepackage{multirow} 
\usepackage{tabularx}
\newcolumntype{C}{>{\centering\arraybackslash}X}

\usepackage{cite}
\usepackage{url}

\usepackage{ulem}
\usepackage{comment} 
\usepackage{color}
\usepackage{ascmac}
\usepackage{fancybox}
\usepackage{seqsplit} 

\title{Clustering algorithm for formations in football games}

\usepackage{authblk}
\author[1]{Takuma Narizuka\thanks{{\it E-mail address}: pararel@gmail.com (T. Narizuka).\\
}}
\author[2]{Yoshihiro Yamazaki}
\affil[1]{Department of Physics, Faculty of Science and Engineering, Chuo University, Bunkyo, Tokyo 112-8551, Japan}
\affil[2]{Department of Physics, School of Advanced Science and Engineering, Waseda University, Shinjuku, Tokyo 169-8555, Japan}
\date{}

\begin{document}
	\maketitle
\begin{abstract}
This paper develops a clustering algorithm for formations in team sports, with a focus on football games.
Our method first clusters formations into several average formations: ``442,'' ``4141,'' ``433,'' ``541,'' and ``343.''
Then, each average formation is further divided into more specific patterns in which the configurations of players are slightly different.
The latter step is based on hierarchical clustering and the Delaunay method, which defines the formation of a team as an adjacency matrix of Delaunay triangulation.
A formation clustered using our method is expressed in a form such as ``442-C1''.
\end{abstract}

\baselineskip 14pt

\section{Introduction}
In competitive team sports, such as football and basketball, each player coordinates with team members and interacts with opposing players.
Throughout such interactions, players maintain a certain formation at the team level.
Such a formation structure reflects a team's strategies for achieving effective attacks and defense in order to win \cite{Hirotsu2009, Tamura2015, Memmert2017, Sumpter2017}.
A traditional method of characterizing formations employs notation such as ``4-4-2,'' which indicates four defenders, four midfielders, and two forwards.
Although this is a convenient means of roughly grasping formation structures, it is too simple to analyze real games.
In fact, the following more quantitative methods have been introduced.

The first example is based on a Voronoi region defined for each player, which is the set of field locations whose distances from the player are less than from any other \cite{Okabe2000}.
Intuitively, this corresponds to the territory of the player on the field.
The basic properties of the Voronoi region have been investigated for football and hockey games \cite{Kim2004, Fonseca2012}, and modified version considering the velocity and acceleration of a player have also been proposed \cite{Taki1996, Taki2000, Fujimura2005, Gudmundsson2014, Gudmundsson2017}.
Bialkowski et al. developed another approach to formations, called ``role representation'' \cite{Bialkowski2014b, Bialkowski2016}. %
Here, the ``role'' represents the relative position of each player in the formation, such as ``center forward'' or ``left wing.''
The key idea behind their method is that players are not distinguished by their uniform numbers, but rather by the role numbers assigned to them.
Specifically, an entire heat map of players' positions for a team is expressed by the sum of 10 heat maps corresponding to roles, in order to yield a minimal overlap between them.
Then, the set of roles is regarded as the formation of the team.
While the previous notation such as ``4-4-2'' is static, the role representation enables more dynamical characterization of formations, e.g., exchange of players' roles during a game.
Along with these studies, we have proposed the Delaunay method, which identifies the adjacency relationships of players' Voronoi regions, i.e., the Delaunay network, with the formation of a team \cite{Narizuka2017}.
Because the formation at time $ t $ is quantified by an adjacency matrix in this method, dissimilarity measures between two different formations can be defined.
On the basis of the Delaunay method, we have also proposed a clustering algorithm for formations in a single game.
This algorithm divides Delaunay networks, which are given at every unit time in a single game, into clusters by means of hierarchical clustering.
We have demonstrated that our method can characterize the differences and dynamics of football formations at different time resolutions within a game by controlling the number of clusters.
The Delaunay method is useful for the time-series analysis and quantitative comparison of formations.
However, the above clustering algorithm for a single game cannot be straightforwardly extended to the case of multiple games.
In this paper, we propose an extended algorithm that can cluster formations over multiple games.

\section{Method}
In the following analysis, we employ datasets comprising 45 football games by 18 teams of J1 League second stage 2016, provided by DataStadium Inc., Japan.
The list of names of 18 teams is as follows:
\begin{quotation}
	``Fukuoka,'' ``Hiroshima,'' ``Iwata,'' ``Kashima,'' ``Kashiwa,'' ``Kawasaki,'' ``Kobe,'' ``Kofu,'' ``Nagoya,'' ``Niigata,'' ``Omiya,'' ``Osaka,'' ``Sendai,'' ``Shonan,'' ``Tokyo,'' ``Tosu,'' ``Urawa,'' ``Yokohama.''
\end{quotation}
There are five games per team and each of five games was taken place on Sept. 25, Oct. 1, Oct. 22, Oct. 29, and Nov. 3.
In this paper, we refer to each game in a form such as Sept. 25 game of ``Sendai.''
Each dataset contains all player positions every 0.04 seconds. 
For simplicity, we focus on the 10 players ($ N=10 $) other than the goalkeeper for each team and analyze the data of the first halves of games where player substitutions did not occur.
We performed the following analysis using python packages; for the calculation of Voronoi region and Delaunay triangulation, {\it Voronoi} and {\it Delaunay} classes in the {\it \seqsplit{scipy.spatial}} module was used; for the hierarchical clustering, {\it linkage} class in the {\it\seqsplit{scipy.cluster.hierarchy}} module was used.
All calculations were executed on a MacBook pro with a 2 GHz Intel Core i5 processor and 16 GB of memory.
The position of the $ j $-th player of a team at time $ t $ is denoted as $ \vec{r}_{j}(t) = [x_{j}(t),\ y_{j}(t)] $.
The centroid position and standard deviation of a team respectively defined as follows:
\begin{align}
	\vec{r}_{c}(t) 
	&= \frac{1}{N}\sum_{j=1}^{N} \vec{r}_{j}(t), \\[10pt]
	\sigma(t) &= \sqrt{\frac{1}{N} \sum_{j=1}^{N}|\vec{r}_{c}(t) - \vec{r}_{j}(t)|^{2}}.
\end{align}
Using $ \vec{r}_{c}(t) $ and $ \sigma(t) $, the normalized coordinates $ \vec{R}_{j}(t) $ for the $ j $-th player are calculated as
\begin{align}
	\vec{R}_{j}(t) &= \frac{\vec{r}_{j}(t) - \vec{r}_{c}(t)}{\sigma(t)}.
	\label{normalize}
\end{align}
\section{Clustering algorithm for a single game}
\label{sec:alg_single}
\begin{figure}[H]
	\centering
	\includegraphics[width=12cm]{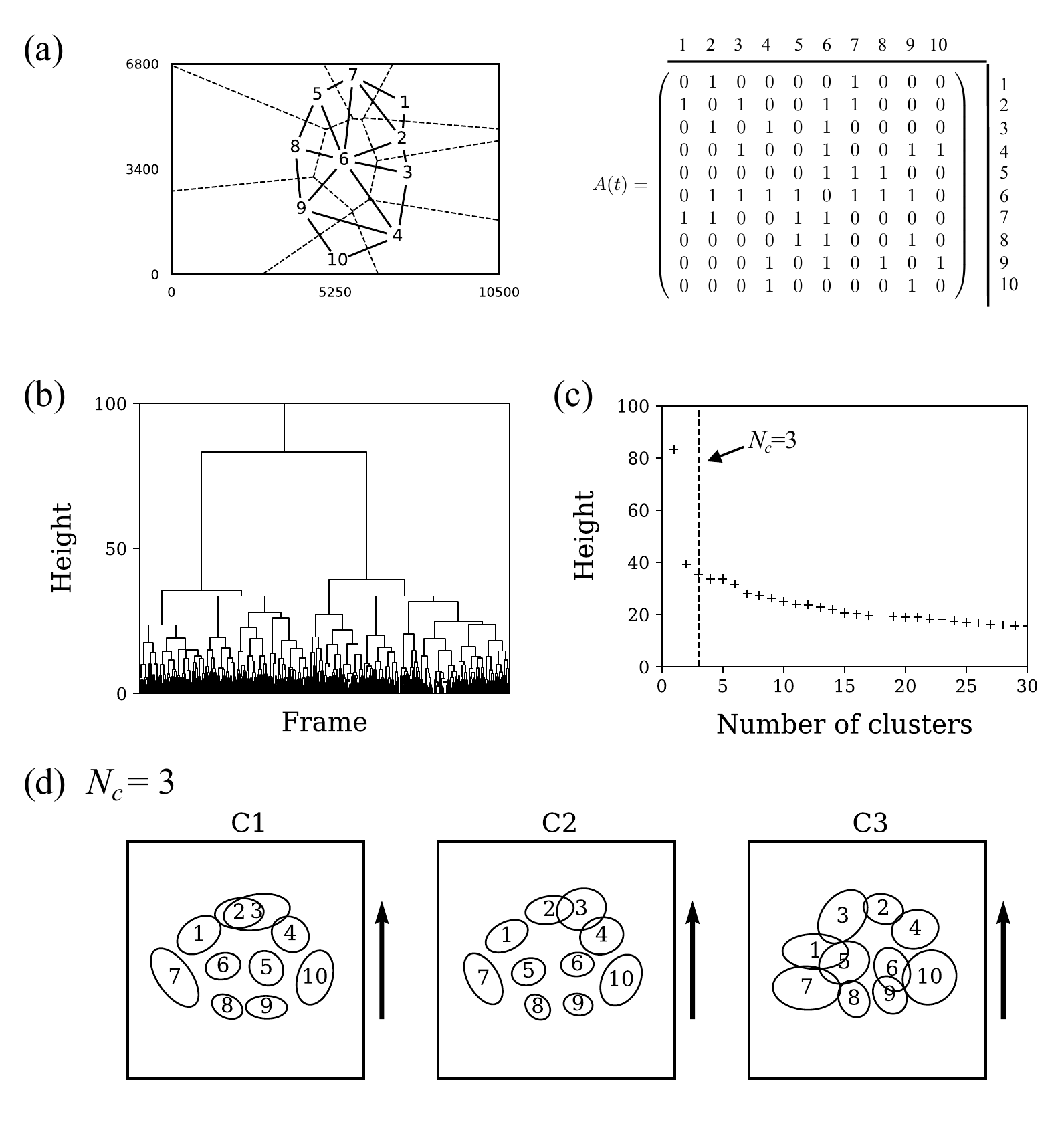}
	\caption{A typical example of a clustering process for Sept. 25 game of ``Sendai.'' (a) A Delaunay network at a certain frame and its adjacency matrix. (b) The Dendrogram and (c) the relation between the number of clusters and height, which are obtained from the hierarchical clustering. The height of the vertical axes corresponds to the distance between two merged clusters. (d) Coarse-grained formations where $ N_{c}=3 $. The major difference between clusters is that players 2 and 3, and players 5 and 6 exchange their positions.}
	\label{fig:clus_single}
\end{figure}

Here, we summarize the Delaunay method and the clustering algorithm for a single team \cite{Narizuka2017}.
In our method, we regard a football formation as adjacency relationships of players, which is independent of the deviation $ \sigma(t) $ of the team.
Specifically, as shown in Fig. \ref{fig:clus_single}(a), a formation of a team at time $ t $ is quantified using the adjacency matrix $ A(t) $ of the Delaunay network, whose components $ A_{ij}(t) $ are given by
\begin{align*}
	A_{ij}(t) &= 
	\begin{cases}
	1 & \textrm{if the Voronoi regions of players $ i $ and $ j $ are adjacent with each other at $ t $,}\\
	0 & \textrm{otherwise}.
	\end{cases}
\end{align*}
Although there are other options for the definition of neighbors in 2D space, we choose the Delaunay triangulation since it is reasonable for the visualization and clustering of formations as shown below.
Owing to this quantification, a dissimilarity measure between two formations at different times can be introduced as
\begin{align}
	D_{tt'} 
	&= \lVert A(t) - A(t') \rVert^{2}
	 = \sum_{i=1}^{N} \sum_{j=1}^{N} [A_{ij}(t) - A_{ij}(t')]^{2}.
	\label{eq:dissim}
\end{align}
Here, we define $ D_{tt'} $ as the Euclidean squared distance, considering the hierarchical clustering using Wards' method.
The dissimilarity $ D_{tt'} $ becomes large when a number of edges are rewired due to the positional exchange of players within the formations.

Based on this dissimilarity measure, we introduced a clustering algorithm for formations appearing in a single game through the following four steps (i)-(iv).
(i) The Delaunay networks every $ \Delta f $ frames in a single game are computed.
(ii) Hierarchical clustering is performed using Ward's method \cite{Tan2006}, where
the input to the clustering is the dissimilarity matrix $ D $ whose components are $ D_{tt'} $ defined by Eq. \eqref{eq:dissim}.
In the Wards' method, distance between two clusters $ C_{1} $ and $ C_{2} $ is given by
\begin{align}
	h(C_{1}, C_{2})
	&= \frac{2n_{1}n_{2}}{n_{1}+n_{2}}\left\lVert \frac{1}{n_{1}} \sum_{t_{1} \in C_{1}} A(t_{1}) - \frac{1}{n_{2}} \sum_{t_{2} \in C_{2}} A(t_{2})\right\rVert^{2},
	\label{ward}
\end{align}
where $ n $ represents the size of $ C $.
From Eq. \eqref{ward}, $ h(C_{1}, C_{2}) $ equals to Eq. \eqref{eq:dissim} at the initial state where each cluster contains one Delaunay network.
In addition, Ward's method is likely to yield comparable size of clusters at each hierarchy compared with other methods.
(iii) The clustering process in step (ii) is displayed by the dendrogram whose vertical axis (height) corresponds to $ h(C_{1}, C_{2}) $ between two merged clusters $ C_{1} $ and $ C_{2} $. 
Particular number $ N_{c} $ of clusters are extracted by cutting the dendrogram at a certain height $ h_{c} $.
(iv) Coarse-grained formations are visualized from each cluster as follows.
For each Delaunay network in a cluster, the positional coordinates of each player are converted into normalized coordinates using Eq. \eqref{normalize}.
This transformation enables to compare each Delaunay network independently of $ \vec{r}_{c} $ and $ \sigma(t) $.
Next, the time averaged position of each player is visualized by an ellipse whose direction and magnitude are determined by the eigenvector and eigenvalue of covariance matrix of each player's position.
The use of the hierarchical clustering enables to control the number of clusters $ N_{c} $ according to the resolution of formations; in fact, if we want to characterize the formation changes in a short time interval, large $ N_{c} $ is selected and vice versa.
As an example, we demonstrate the above clustering process based on Sept. 25 game of ``Sendai.''
Figure \ref{fig:clus_single}(b) is the dendrogram obtained from the step (ii) where $ \Delta f = 25 $.
The optimal number of clusters $ N_{c} $ can roughly be determined as the point where height increases rapidly with decreasing of the number of clusters (Fig. \ref{fig:clus_single}(c)); in this case, we chose $ N_{c}=3 $.
We show the clusters (coarse-grained formations) where $ N_{c}=3 $ in Fig. \ref{fig:clus_single}(d).
Each cluster is distinguished by a cluster number from C1 to C3, and the difference between them is that several pairs of players exchange their positions: players 2 and 3, and players 5 and 6, for example.
\section{Clustering algorithm for multiple games}
\label{sec:alg_multi}
Let us consider the problem of clustering Delaunay networks over multiple games.
In order to quantify a formation using an adjacency matrix $ A(t) $, uniform numbers $ \vec{U}=[a,b,\ldots, j] $ of players need to be assigned to the indexes $ \vec{I} = [1,2,\ldots, 10] $ of $ A(t) $.
If we cluster Delaunay networks of a single game in which no player substitutions occur, then an arbitrary correspondence between $ \vec{U} $ and $ \vec{I} $ can be employed.
However, clustering over multiple games requires the assignment of multiple uniform numbers $ \vec{U}_{1}, \vec{U}_{2}, \ldots $ for different games to one set of indexes $ \vec{I} $.
Because such an assignment is not uniquely determined, a criterion must be defined. 
Here, we adopt the framework of ``role representation'' introduced in \cite{Bialkowski2014b, Bialkowski2016}.
We assume that the players play the same roles if they occupy similar positions in a formation.
Then, we label each player by a role number and identify them with the indexes $ \vec{I} $ of $ A(t) $.
In the following, we propose an extended clustering algorithm based on this idea, consisting of three parts I, II, and III.
\subsection*{Part I: clustering into average formation}
\begin{figure}[H]
	\centering
	\includegraphics[width=16cm]{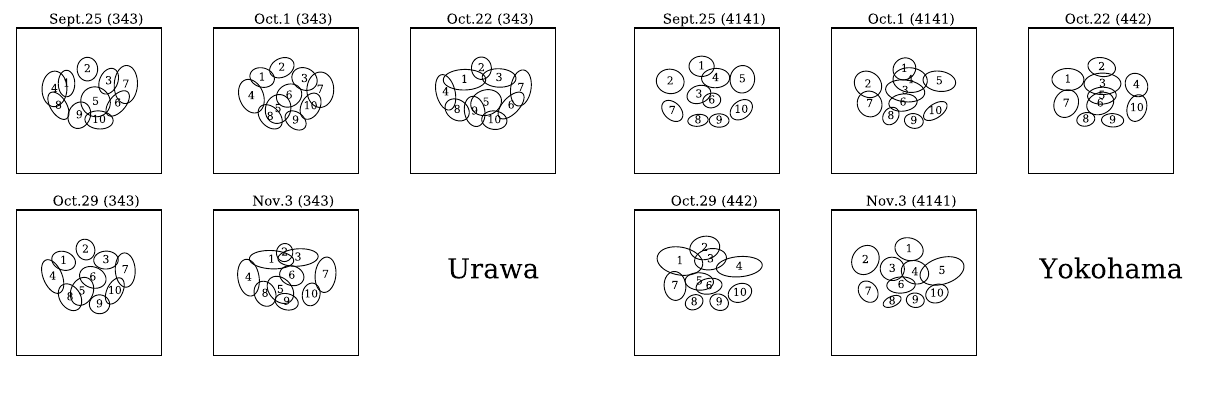}
	\caption{Heat maps (average formations) of five games for 18 teams. The direction of offense is upward in each panel. Each heat map belongs to one of the following five average formations: ``442,'' ``4141,'' ``433,'' ``541,'' and ``343.'' The label ``(others)'' means that player substitutions occurred in the first half of the game, or the average formation could not be identified.}
	\label{fig:ave_form_team}
\end{figure}

\begin{figure}[H]
	\centering
	\includegraphics[width=16cm]{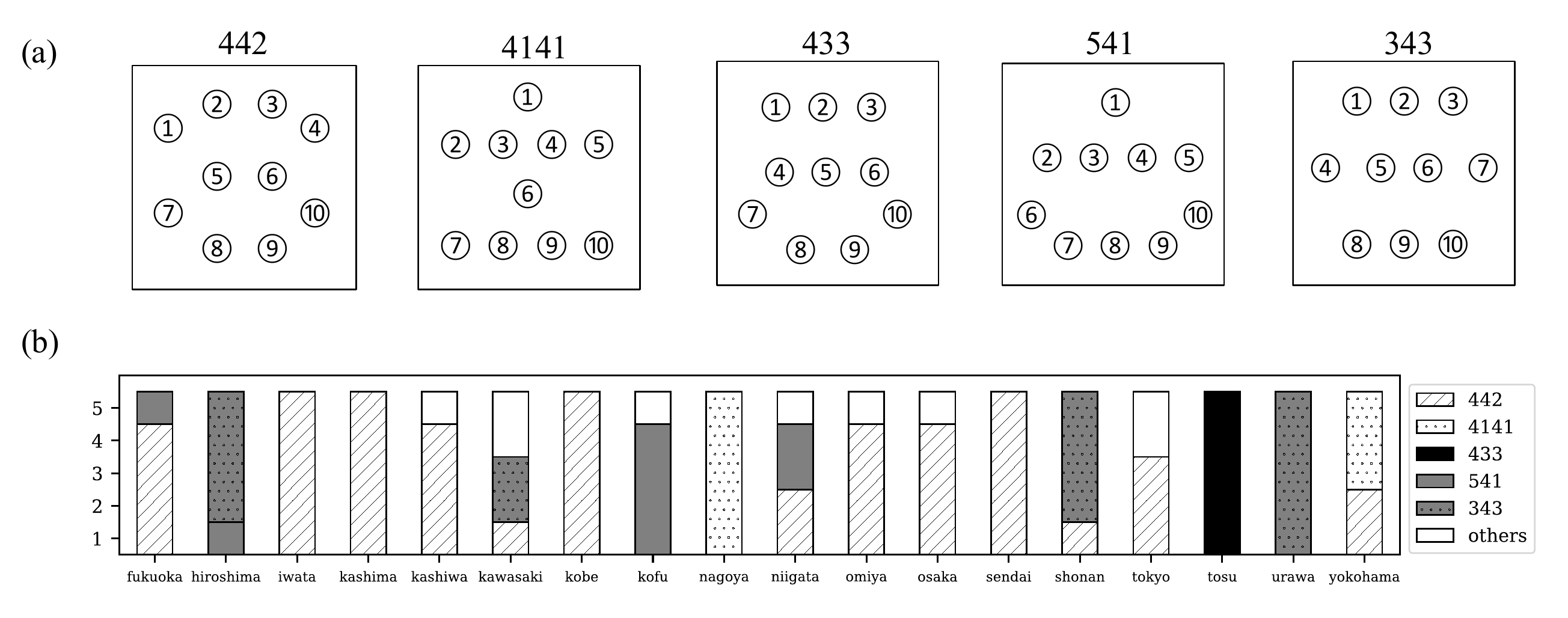}
	\caption{(a) Schematic representation of each average formation. The heat map of each team shown in Fig. \ref{fig:ave_form_team} belongs to one of these five patterns. (b) Average formations of each team throughout five games. The label ``others'' means that player substitutions occurred in the first half of the game, or the average formation could not be identified.}
	\label{fig:ave_form}
\end{figure}

In part I, we assign the same index $ i $ of $ A(t) $ to players whose positions in a formation are approximately the same.
To estimate the position of each player in a game, we compute the heat map of each game for each team in the normalized coordinates \eqref{normalize}.
We present the heat maps obtained for all teams and games in Fig. \ref{fig:ave_form_team}.
In this figure, the position of each player is expressed by the region within each ellipse.
The direction and magnitude of an ellipse are determined by the eigenvector and eigenvalue of the covariance matrix for the corresponding player's position.
These heat maps appear to be classified into several patterns.
In fact, we find from our data that they belong to one of the following five patterns: ``442,'' ``4141,'' ``433,'' ``541,'' and ``343'' (these are referred to as ``average formations'' hereafter).
A schematic representation of the five average formations is shown in Fig. \ref{fig:ave_form}(a).
The frequency of such formations for each team in five games is shown in Fig. \ref{fig:ave_form}(b).
It should be noted that we manually classified the heat maps into the average formations.
Almost all teams, except the teams with player substitutions in the first half, can be classified into one of the average formations.
Hence, the change in average formations during a game did not occur in our data.
We also note that the names of the average formations are not an official one and other notations can also be considered.

For a certain average formation, the ellipses (average positions of players in a game) are distinguished by serial numbers from 1 to 10, as shown in Fig. \ref{fig:ave_form}(a).
It is considered that players belonging to the same average formation with the same serial number play the same role in the team (e.g., player 1 in ``4141'' is interpreted as a ``center forward'').
Therefore, we identify these serial numbers with the indexes $ \vec{I} $ of $ A(t) $, and a one-to-one correspondence between $ \vec{U}_{1}, \vec{U}_{2}, \ldots $ and $ \vec{I} $ is obtained for each average formation.
\begin{figure}[H]
	\centering
	\includegraphics[width=16cm]{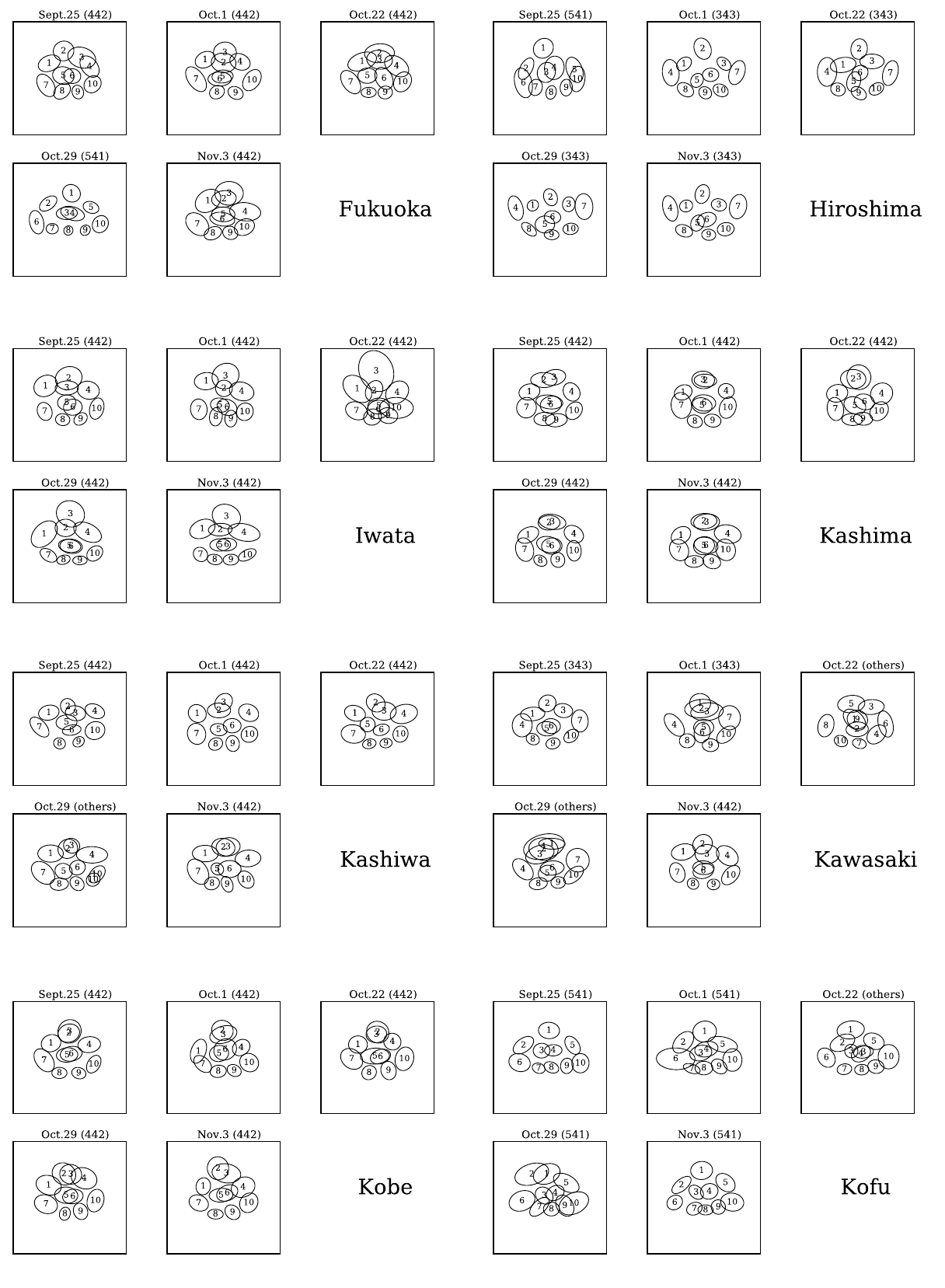}
\end{figure}
\begin{figure}[H]
	\centering
	\includegraphics[width=16cm]{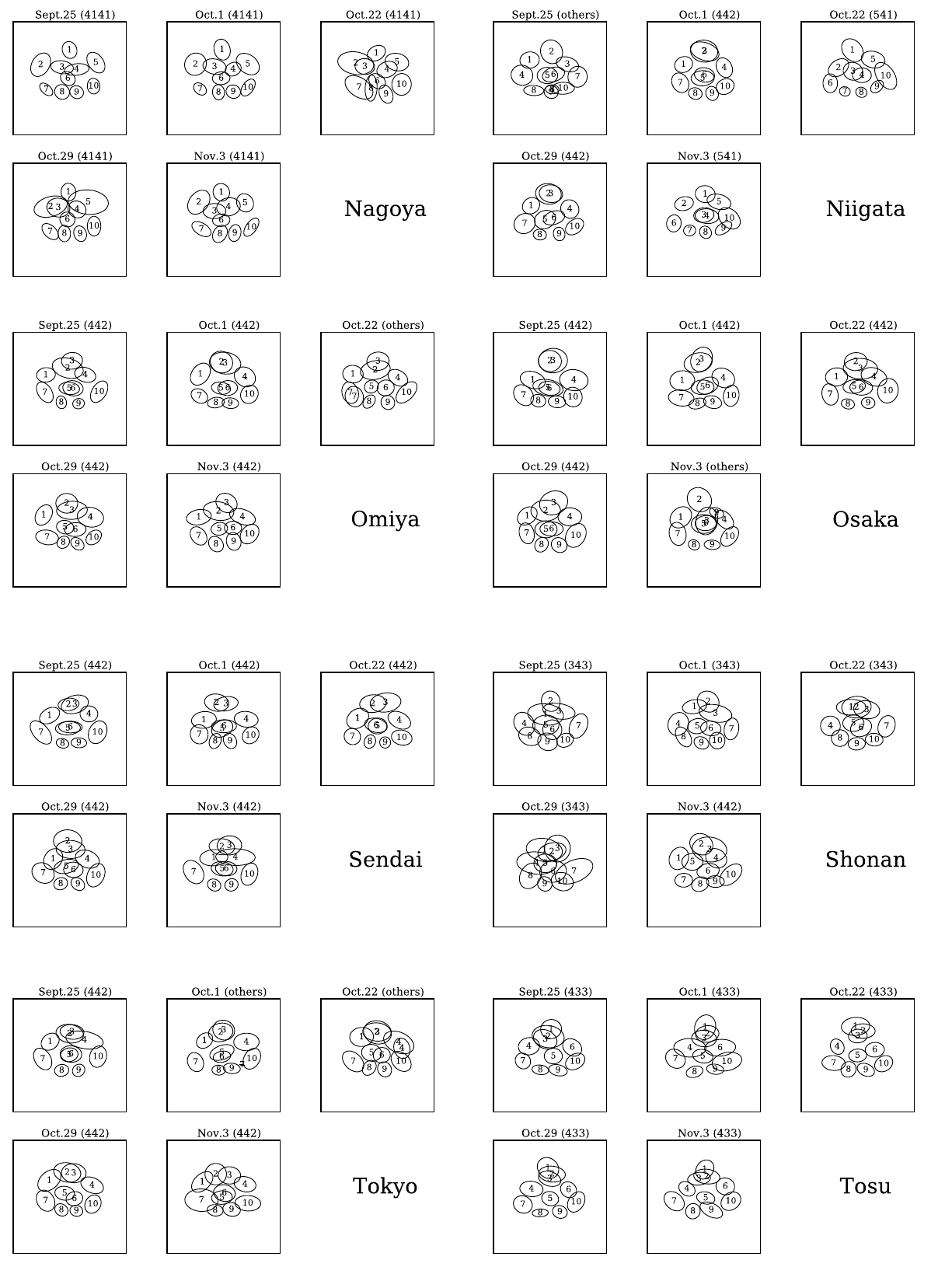}
\end{figure}

\subsection*{Part II: hierarchical clustering of average formations}
\begin{figure}[H]
	\centering
	\includegraphics[width=12cm]{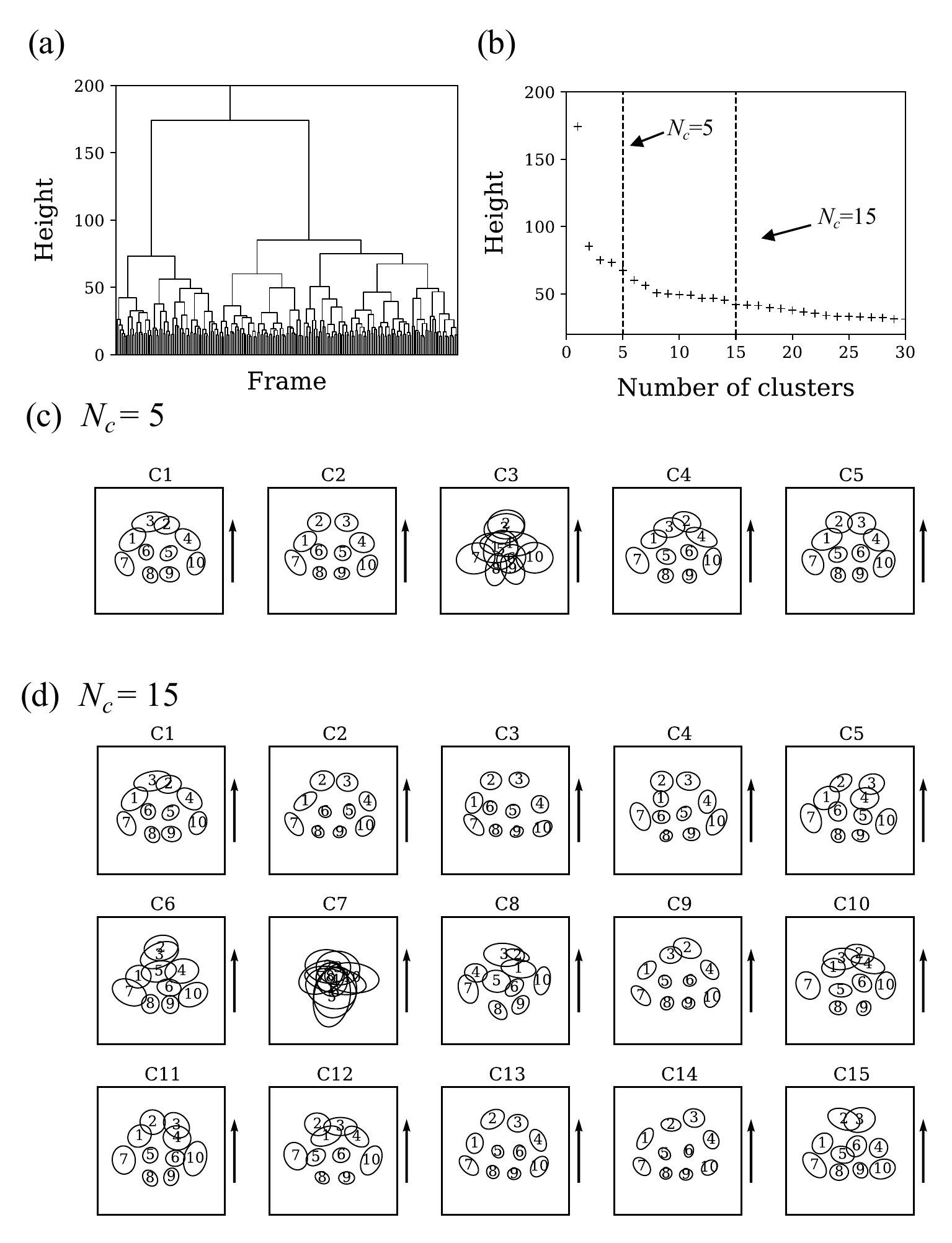}
	\caption{Results of hierarchical clustering for the five games of ``Sendai.'' (a) The Dendrogram, (b) the relation between the number of clusters and height, and the visualization of coarse-grained formations where (c) $ N_{c}=5 $, and (d) $ N_{c} = 15 $. Each cluster is distinguished by a cluster number.}
	\label{fig:clus_multi}
\end{figure}

As shown in Fig. \ref{fig:ave_form_team}, the ellipses of some players in a heat map overlap, indicating that these players exchange their positions or move close to each other in the game. 
In addition, the configurations of players are slightly different even within the same average formation.
In order to distinguish such patterns, in part II, we cluster all the Delaunay networks belonging to the same average formation using the clustering algorithm introduced in Sec. \ref{sec:alg_single}. 
Figure \ref{fig:clus_multi} presents typical examples of clustering results for the five games of ``Sendai'' where $ \Delta f = 25 $, with $ N_{c} = $ 5 or 15. 
Because ``Sendai'' adopted ``442'' in all five games (see Fig.\ref{fig:ave_form}(b)), the coarse-grained formation obtained using this method is expressed as ``442-C1,'' where the former number denotes the average formation and the latter is the cluster number.
Furthermore, each ellipse in a cluster in Fig. \ref{fig:clus_multi} consists of all the positions of players with the same index in the five teams.
We find that each cluster exhibits a more specific pattern compared with the corresponding average formations.
The major difference between clusters is that players 2 and 3, or players 5 and 6 exchange their positions.
We note that C3 in $ N_{c}=5 $ or C7 in $ N_{c}=15 $ only include irregular patterns, which could be associated with transitional situations such as competition in front of goal or counter attacks.
The value of $ N_{c} $ depends on the cutting height $ h_{c} $ of the dendrogram, where the height represents the distance between two merged clusters in the clustering process.
As noted in Sec. \ref{sec:alg_single}, we can control the degree of coarse-graining of formations by varying $ N_{c} $: finer (coarser) patterns are obtained by increasing (decreasing) $ N_{c} $.
For example, C2 in $ N_{c}=5 $ is divided into (C2, C3, C4, C5) in $ N_{c}=15 $; C5 in $ N_{c}=5 $ is divided into (C11, C12, C13, C14, C15) in $ N_{c}=15 $.
In addition, when $ N_{c}=15 $, the positions of players 7 and 10 are slightly different between clusters compared with the case of $ N_{c}=5 $.
In particular, there are two patterns that players 7 and 10 are in a middle line and a back line; such two patterns appear to correspond to the offense and defense scenes, respectively.
We note that the special case, $ N_{c}=1 $, is the most coarse pattern, corresponding to the superposition of all average formations of the five games.

\subsubsection*{Part III: transition network between clusters}
\begin{figure}[H]
	\centering
	\includegraphics[width=16.5cm]{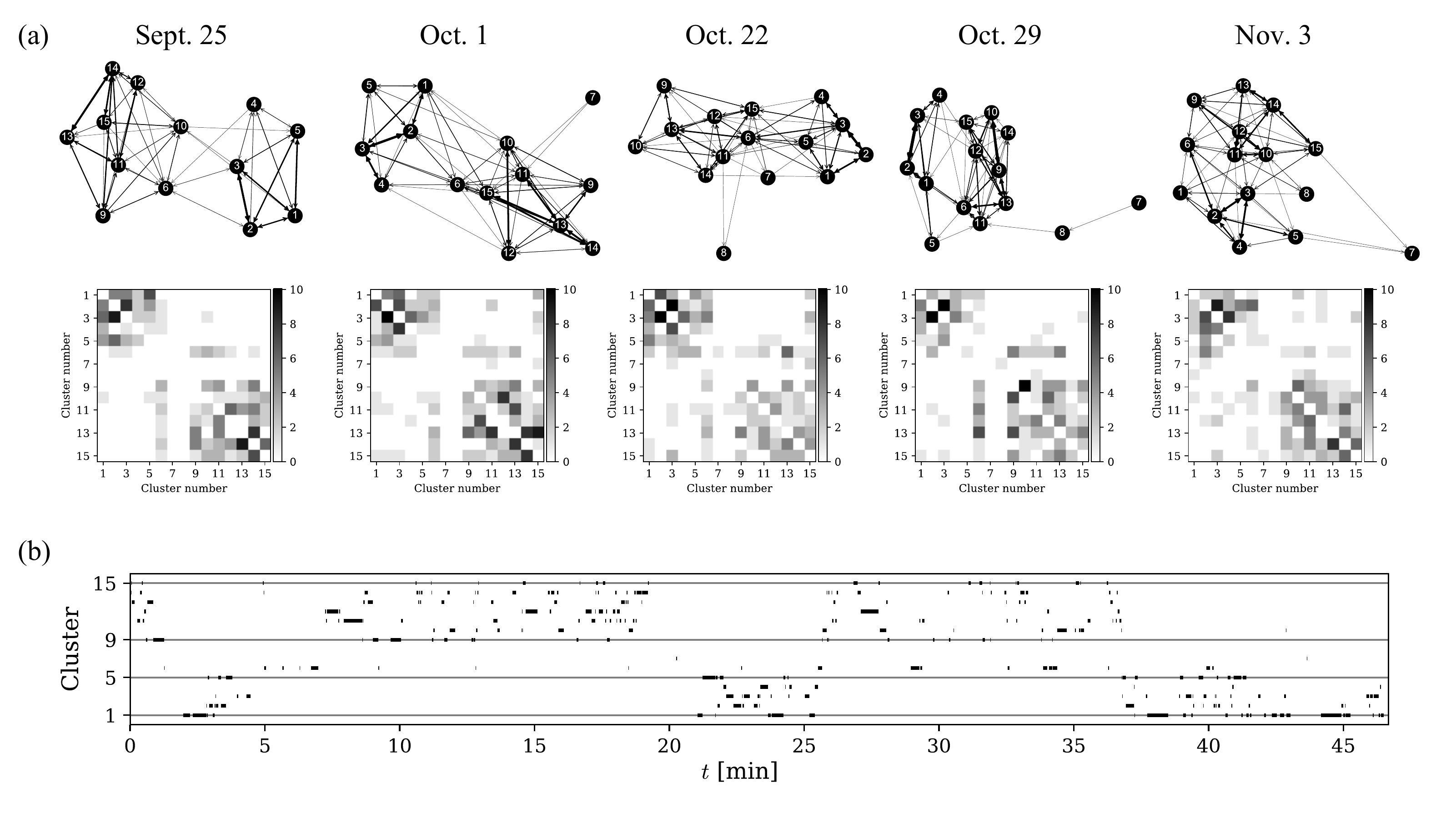}
	\caption{(a) Transition networks between clusters (upper panels) and those of adjacency matrices (below panels) for all games of ``Sendai'' where $ N_{c}=15 $. Each node corresponding to the coarse-grained formation in Fig. \ref{fig:clus_multi}(d) is arranged using Fruchterman-Reingold force-directed algorithm \cite{Fruchterman1991}. (b) Time series of the clusters for Sept. 25 game of ``Sendai.''}
	\label{fig:transition}
\end{figure}

When a certain number $ N_c $ of clusters is given, a continuous time series of formation changes can be regarded as discrete transitions between the clusters.
In Fig. \ref{fig:transition}(a), we present transition networks, whose nodes and edges represent clusters and number of transitions between them, for the five games of ``Sendai.''
Here, each node in the networks corresponds to the coarse-grained formation shown in Fig. \ref{fig:clus_multi}(d), and a transition from one cluster to another represents a change of the configuration of players in the formation; e.g., C1 $ \to $ C2 indicates that the players 2 and 3 exchange their positions.
The nodes are placed using Fruchterman-Reingold force-directed algorithm \cite{Fruchterman1991}, which achieves an optimal layout depending on the number of transitions between clusters (weight of edges): two nodes with a large number of transitions are placed in nearby locations.
In addition, we also visualize adjacency matrices of corresponding transition networks in Fig. \ref{fig:transition}(a).
We find from Fig. \ref{fig:transition} that each of the five games exhibits similar transition patterns as follows.
First, there are two communities consisting of clusters (C1, C2, C3, C4, C5), and (C9, C10, C11, C12, C13, C14, C15); the former (latter) community corresponds to the pattern that the player 5 is on the right (left) and the player 6 is on the left (right).
Second, cluster C6 is the coarse-grained formation connecting such two communities; in fact, players 5 and 6 are lined up vertically in the formation.
Third, clusters C7 and C8 are somewhat irregular formations, e.g., positions of players 1 and 4 in C8 are different from other clusters.
It is noted that each community includes a cluster corresponding to the position-exchanged pattern between players 2 and 3, i.e., C1 and (C9, C10).
We further show the time series of the clusters for Sept. 25 game of ``Sendai'' in Fig. \ref{fig:transition}(b).
We find that the transition between two communities occurs only a few times in the first half; namely, if players 5 and 6 exchange their positions once, the formation continues for a while. 
On the other hand, the duration time of the clusters C1 and (C9, C10) is not such a long, and players 2 and 3 exchange positions more frequently.
Because we confirmed that such features are in common for the five games, this appears to regard the strategy of ``Sendai.''

\section{Discussion}
We have proposed an extended clustering algorithm based on role representation (part I) and hierarchical clustering (part II).
Here, we compare our clustering algorithm with the method introduced by Bialkowski et al. \cite{Bialkowski2014b, Bialkowski2016}.
In that method, a 2D probability distribution $ H(\vec{r}) $ (heat map) for a team is divided into 10 heat maps, $  H(\vec{r}) = \sum_{r=1}^{10} H_{r}(\vec{r}) $, and the set $ {\cal F} = \{H_{r}(\vec{r}); r=1, \cdots, 10 \} $ is regarded as the formation.
Each $ H_{r}(\vec{r}) $ is computed to achieve a minimal overlap with others, under the condition that each player belongs to a different $ r $ at each frame.
Because each player is labeled by a role number $ r $ instead of uniform number $ u $ at each frame, this method is called ``role representation.''
In the role representation approach, $ H_{r}(\vec{r}) $ consists of various players at different frames, and patterns in which two players exchange their positions are regarded as the same.
In contrast, our algorithm describes an entire heat map $ H(\vec{r}) $ as the sum of players' heat maps, $ H(\vec{r}) = \sum_{u=1}^{10} H_{u}(\vec{r}) $, where $ u $ denotes the uniform number.
The set $ {\cal F} = \{H_{u}(\vec{r}); u=1, \cdots, 10 \} $ is called a ``average formation.''
This decomposition does not achieve the minimal overlap, namely, players with different $ u $ can exchange their positions during a game.
Instead, our method distinguishes such position-exchanged patterns as different formations, based on the Delaunay method and hierarchical clustering; in particular, the quantification of a formation as the Delaunay triangulation is essential because it can incorporate the information of adjacency relationships of players.
In this sense, our method realizes a more detailed characterization of formations compared with that in \cite{Bialkowski2014b, Bialkowski2016}, although we have only shown the results for the particular datasets.
While our decomposition of the entire heat map $ H(\vec{r}) $ does not achieve the minimal overlap, the average positions of players, expressed by ellipses, are still clearly separated (see  Fig. \ref{fig:ave_form_team}).
That is, each player carries out an individual role in a football game.
This feature of football games allows us to label players not only by uniform numbers $ \vec{U} $ but also by role numbers (role representation).
Furthermore, it provides a criterion for the correspondence between multiple uniform numbers $ \vec{U}_{1}, \vec{U}_{2}, \ldots $ and the indexes $ \vec{I} $ of $ A(t) $, and allows hierarchical clustering to be realized over multiple games.
We note that our method can be applied to specific sports in which players' average positions are almost fixed because it relies on the one-to-one correspondence between $ \vec{U} $ and $ \vec{I} $.
The variation in average formations and switches among them are a reflection of teams' strategies \cite{Hirotsu2009, Tamura2015}.
It has been reported that football teams adopt a so-called ``win-stay lose-shift strategy'' for formation changes between games \cite{Tamura2015}: they tend to adopt the same (a different) formation after a win (loss).
Our method has the potential to provide a more detailed characterization of strategies or game flow by focusing on formation changes within a game.
As an example, we have introduced the transition networks between clusters in Fig. \ref{fig:transition}.
While we have mentioned some common features in Sec. \ref{sec:alg_multi}, a closer look at the adjacency matrices in Fig. \ref{fig:transition} shows that each network exhibits slightly different transition patterns.
In order to extract more specific patterns from them, more large $ N_{c} $ is needed.
We expect that temporal network analysis for the cluster transitions for different $ N_{c} $ values to provide insights into the characterization of team styles.
Regarding this type of analysis, a further extension of the Delaunay network could also be considered.
In fact, the present Delaunay network lacks information on opposing teams.
This means that the edges do not always represent pass routes, because opposing players may exist on these edges.
We can address this problem by introducing a Delaunay triangulation method including an opposing team.
Further dynamical analyses of formation structures incorporating ball passes or interactions with opposing players by employing this extended Delaunay network will be a topic of future research.
Finally, the Delaunay method and the clustering algorithm using hierarchical clustering are a general framework to coarse grain a many-particle system with incorporating its adjacency relationships.
It realizes more detailed characterization and visualization rather than macroscopic quantities such as the centroid and the standard deviation for collective motions of various systems, including team sports \cite{Gudmundsson2014}, animals \cite{Sumpter2010}, and robots \cite{Deblais2018}.
We expect that our method will provide a common tool for formation analysis of team sports and new insights to the research fields of general collective motions.

\section{Conclusion}
We have proposed a new algorithm that can cluster formations in team sports over multiple games.
A formation in our method is expressed by the combination of an average formation and a cluster number, such as ``442-C1.''
The Delaunay method and hierarchical clustering divide the average formation into more specific patterns (clusters) that several players change their relative positions within a formation.
Applying our algorithm to the datasets comprising football games of J1 League, we have demonstrated the formation analysis.
Based on the transition network between clusters, we have extracted typical transition patterns of the formation for a particular team.
We conclude that our method provides a common tool for formation analysis of various systems and is helpful to characterize team styles in team sports.

\section*{Acknowledgements}
The authors are very grateful to DataStadium Inc., Japan for providing the player tracking data. 
The authors thank H. Kuninaka and T. Mizuguchi for a fruitful discussion.
This work was partially supported by the Data Centric Science Research Commons Project of the Research Organization of Information and Systems, Japan, a Grant-in-Aid for Young Scientists 18K18013 from the Japan Society for the Promotion of Science (JSPS), and Hayao Nakayama Foundation for Science, Technology and Culture.
%

\bibliography{./reference}

\end{document}